\documentclass[prb,aps,reprint]{revtex4-1}
\usepackage{graphicx}
\usepackage[usenames,dvipsnames]{color}
\usepackage[normalem]{ulem}
\usepackage{amsmath}
\usepackage{amsfonts}
\usepackage{amssymb}
\usepackage{fmtcount}

\newcommand{\pav}{p_\text{av}}
\newcommand{\pbond}{p_\text{bond}}
\newcommand{\pc}{p_\text{c}}
\newcommand{\pcon}{p_\text{con}}

\begin{document}
\title{Percolation thresholds for discrete--continuous models with non-uniform probabilities of bond formation} 
\author{Bart\l{}omiej Szczygie\l{}}
\email[]{bartlomiej.szczygiel@students.mimuw.edu.pl}
\affiliation{College of Inter-Faculty Individual Studies in
Mathematics and Natural Sciences, University of Warsaw, \.Zwirki i
Wigury 93, 02-089 Warsaw, Poland}
\author{Marek Dudy\'{n}ski}
\email[]{marek.dudynski@mtf.pl}
\affiliation{Modern Technologies and Filtration, Przybyszewskiego 73/77 lok. 8, 01-824 Warsaw, Poland }
\author{Kamil Kwiatkowski}
\email[]{kamil.kwiatkowski@fuw.edu.pl}
\altaffiliation{} 
\affiliation{Institute of Theoretical Physics,
Faculty of Physics, University of Warsaw, Pasteura 5, 02-093
Warsaw, Poland}
\affiliation{Interdisciplinary Centre for Mathematical and
Computational Modeling, University of Warsaw, Prosta 69, 00-838
Warsaw, Poland}
\author{ Maciej Lewenstein}
\email[]{ maciej.lewenstein@icfo.es  }
\affiliation{ICFO-Institut de Ci\`encies Fot\`oniques, The Barcelona Institute of Science
and Technology, Av. Carl
Friedrich Gauss 3, 08860 Barcelona, Spain}
\affiliation{ICREA-Instituci\'o Catalana de Recerca i Estudis
Avan\c cats, Lluis Campanys 23, 08010 Barcelona, Spain}
\author{Gerald John Lapeyre Jr}
\email[]{ john.lapeyre@icfo.es }
\affiliation{Spanish National Research Council (IDAEA-CSIC),
E-08034 Barcelona, Spain}
\affiliation{ICFO-Institut de Ci\`encies
Fot\`oniques, The Barcelona Institute of Science
and Technology, Av. Carl Friedrich Gauss 3, 08860 Barcelona, Spain}
\author{Jan Wehr}
\email[]{ wehr@math.arizona.edu}
\affiliation{Department of Mathematics, University of Arizona, Tucson AZ 85721, USA}
\date{\today}
\begin{abstract}
We consider a family of percolation models in which geometry and
connectivity are defined by two independent random processes.
Such models merge characteristics of discrete and continuous
percolation. We develop an algorithm allowing effective
computation of both universal and model-specific percolation
quantities in the case when both random processes are Poisson
processes. The algorithm extends percolation algorithm
by Newman and Ziff (M.E.J. Newman and R.M. Ziff, Phys Rev E, 64(1):016706, 2001) to handle inhomogeneous lattices. In particular, we use the proposed method to compute critical exponents and cluster density distribution in two and
three dimensions for the model of parallel random tubes
connected randomly by bonds, which models the connectivity properties of activated carbon.
\end{abstract}
\maketitle 
\section{Introduction}
Two basic types of percolation models are discrete and continuous
percolation \cite{grimmett2010percolation,Stauffer}. In the
discrete case, a lattice is given and its bonds (edges) are
open, or its sites (vertices) are occupied, with a probability $p$,
which is the relevant parameter of the model. Depending on the
case, we speak of {\it bond percolation} or {\it site
percolation}. The local random variables, which determine bond openness or site
occupations, define global connections and the main focus of the
theory is the phenomenon of {\it percolation}, i.e. the appearance
of an infinite cluster (or, in some models: of infinite clusters)
of connected bonds or sites. 

\begin{figure}[htp]
\includegraphics[width=0.45\columnwidth]{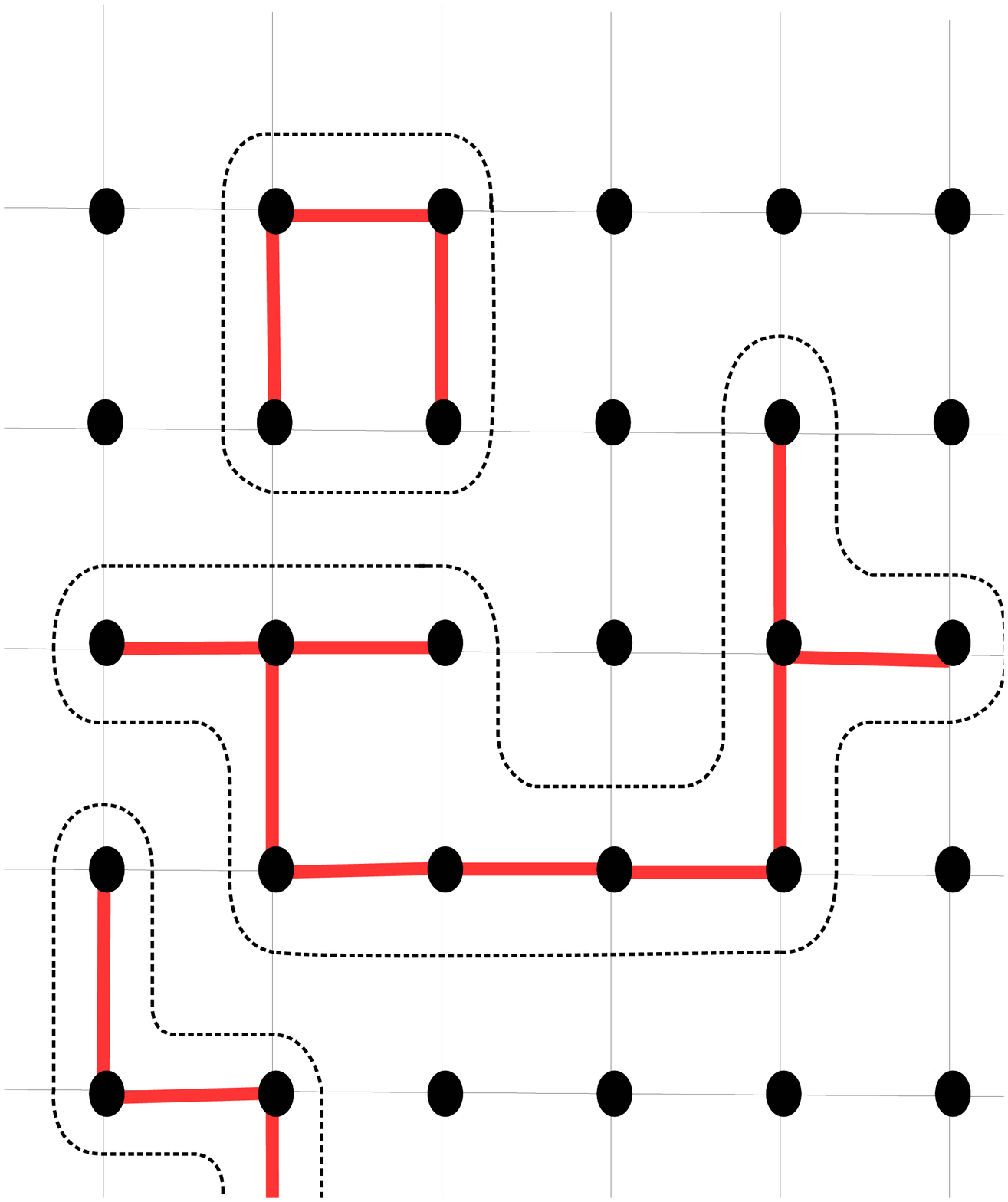}
\includegraphics[width=0.45\columnwidth]{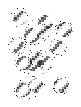} \\
a) discrete percolation \hspace{0.1\columnwidth} b) continuum percolation
        \caption{\label{fig:clasic} Examples of percolation discrete and continuum percolation: a) bond percolation on the square lattice, b) discs in the plane. Clusters are delineated in both cases.
}
 \end{figure}

In continuum models the positions of percolating objects
themselves are chosen at random in space and the connections are
determined solely by the realization of the objects
\cite{meester1996continuum}. A parameter $\eta$ playing a role
analogous to $p$ is usually defined as  the expected value of the
local density of the objects. We will usually refer to $\eta$ or
$p$ as the model parameters. In the discrete approach, one can also
generate the lattice randomly, and then open its edges
with the same probability, independently of the random geometry.
Classical examples of discrete and continuum percolation are
presented in Fig. \ref{fig:clasic}.
However, there are instances when complexities of percolation
phenomena are beyond the scope of these two basic types of percolation model. A
simple example is a system of roads, in which width of a road is describe by the weight of the corresponding edge and the traffic intensity corresponds to the percolation parameter. In this situation the probability of a road connection
between two points being open is a function of both these
parameters \cite{PhysRevLett.96.148702,Li20012015}.
Another interesting case, so-called radio tower model \cite{franceschetti2005continuum}, is obtained by modifying the disc percolation model \cite{Discs_2003}. In this model we first randomly distribute in the plane points (towers) which are the centers of discs with fixed radius $R$. The different towers cannot communicate beyond the distance $R$, which is the parameter of the model. 
We set the probability that a connection (an open bond) exists between a pair of towers as $\pbond = \max(0,1-d/R)$, where $d$ is a distance
between two points. We look for the critical value of $R$ at which
an infinite cluster appears. 

These models have two things in common: their geometry is random and the possible connections in the system are determined by a random
variable, whose distribution is defined by both the geometry and the model parameter. Models of such discrete-continuous class are thus described by a random graph with weighted edges, where the probability of a connection depends both on the percolation parameter, and on the weight of edges, dictated by the geometry of the graph realization.

In the present paper we introduce a class of discrete-continuous percolation models, consisting of parallel random tubes connected randomly by bonds.
These models merge the characteristics of both discrete and
continuous types of percolation, and are proposed to
describe some connectivity properties of activated carbon \cite{furmaniak2014folding,furmaniak2012simulation}.
In order to compute such properties, however, efficient algorithms for discrete-continuous types of percolation models have to be developed. To meet this challenge, in particular to handle inhomogeneous lattices, we extend an efficient percolation algorithm by Newman and Ziff \cite{Ziff2001}.

Motivations for using such an inhomogeneous tube-based model to
simulate percolative properties of activated carbon originate from
wood processing science. In the process of wood gasification the
material is first transformed into charcoal containing 
approximately one-third of its initial mass, and then into various
stages of activated carbon. Finally the structure of the material breaks down making the material 
 collapse into fine dust, which burns into a small amount of ash
\cite{Feng2000,Kwiatkowski2014125}.  Fragmentation, which is a phenomenon closely related to percolation, 
is observed during thermal conversion of charcoal
\cite{Kwiatkowski2014125,Feng2000}. In this process the initial
structure (skeleton) of wood, composed of parallel cylinders,
persists, but the hemicellulose, cellulose and lignin that form walls of
the cylinders are transformed into more carbonic compounds. In
this way, although the initial skeleton persists, the
microscopic structure of its walls becomes much more complex. Fine
micro-porous substructures \cite{marsh2001activated} are formed,
and lead to a rapid increase of internal surface
(specific surface area), when charcoal is transformed into
activated carbon.  Several models were developed to explain the
complicated micro-structures observed in charcoal and activated
carbon \cite{pawlyta2013transmission}. In particular, various forms
of carbon potentially building such micro-structures were
considered: graphene ribbons (model of Jenkins-Kawamura),
fullerenes (Harris model
\cite{harris2003impact,marsh2006activated}), stacked graphite
\cite{jenkins1971structure} or graphene
\cite{pawlyta2013transmission}, carbon onions \cite{chen2006pore}
and nanotubes \cite{harris2013fullerene,furmaniak2014folding,furmaniak2012simulation}.
In the present paper we explore percolation properties of a tube-based
model, representing a nanopipe micro-structure of activated carbon \cite{harris2013fullerene,wang2014three,furmaniak2014folding,furmaniak2012simulation}. 
We assume that the skeleton walls are made of a collection of parallel tubes representing nanopipes of varying lengths. These
nanopipes form an inhomogeneous lattice bound together by
amorphous carbon connections. We assume that during gasification
with CO$_2$ and H$_2$O, the amorphous carbon
is reacting with these gasification agents, and the bonds are removed.
The bond removal leaves more and more nanopipes disconnected, leading
to disintegration of small clusters and, finally, to the
breakdown on the percolating skeleton.
Potential applications of the introduced class of percolation
models and the developed algorithm are beyond this particular tube-based description 
of activated carbon, including also above
mentioned models of road networks and radio towers, as well as other
discrete-continuous percolation systems.  

In Section \ref{Model} we describe a tube-based percolation model.
We then introduce an extension of the well-knows percolation algorithm
by Newman and Ziff \cite{Ziff2001}, which allows us to treat the
inhomogeneity of the lattice inherent in our model (Section
\ref{algorithm}). We validate the extended algorithm in Section
\ref{res2D}, comparing its results with the known exact solutions for two-dimensional
percolation, and use it to obtain new results for the three
dimensional problem in the final Section \ref{sec:3D}.

\section{Percolation model}
\label{Model} 
Here we present the tube-based model. 
To define the model precisely in two-dimensions, we proceed in three steps:
\begin{itemize}
        \item we start from $n$ parallel (vertical, for definiteness) lines of length $L$.
        \item we use $n$ independent Poisson processes with the same parameter $\mu_1$ to divide the lines into segments, called tubes.
        \item we introduce bonds between each pair of adjacent lines and in this manner the connections between tubes are established. The bonds are generated by independent Poisson processes with parameter $\mu_2$.
\end{itemize}

\begin{figure}[htp]
 \includegraphics[width=0.8\columnwidth]{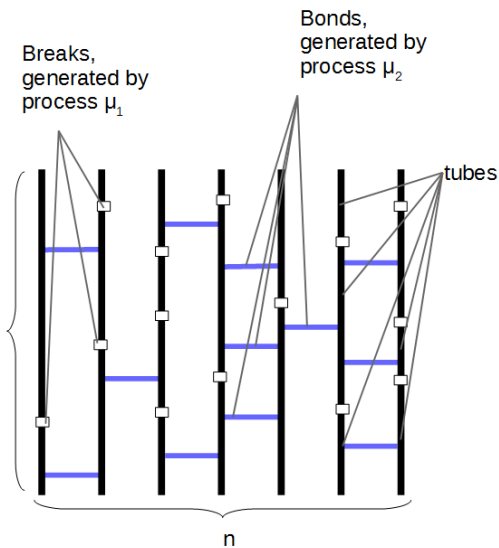}%
 \caption{\label{fig:graph}  An example of a realization of the two dimensional graph described by a set of four parameters $\{L, n, \mu_1, \mu_2\}$.} %
\end{figure}

The resulting graph is presented in Fig. \ref{fig:graph}.

A three dimensional model is defined similarly. First, we
introduce a set of lines of length $L$  passing through the points
of a square lattice and perpendicular to plane of this lattice.
Then we follow the procedure for 2D case, dividing lines into
segments and generating bonds between each pair of adjacent lines.
``Adjacent'' is defined here using the nearest-neighbor
connections on the underlying square lattice, so that in the 2D
case a line not lying on the boundary has two adjacent lines
while in the 3D case it has four. 

The resulting discrete-continuous model consists of parallel tubes of random length connected randomly by bonds whose distribution is defined by the spatial location of the tubes and by the model parameter. 

The resulting random graph model is described by four parameters $\{L, n, \mu_1, \mu_2\}$, defining the size of the model ($L$ and $n$), length of the tubes described by Poisson processes with the parameter $\mu_1$ and with bonds between these tubes generated by independent Poisson processes with the
parameter $\mu_2$.
Under rescaling in the direction of the lines, the resulting graph is equivalent to the system with parameters $\{L\mu_1, n, 1, \mu_2/\mu_1\}$. We thus put $\mu_1:=1$ and
$\mu_2=\mu$, so in the limit when $L$ and $n$ go to infinity at
the same rate the model has only one  parameter $\mu$. For
simplicity in most of the simulations we put $L=n$. 

\begin{figure}[htp]
 \includegraphics[width=0.7\columnwidth]{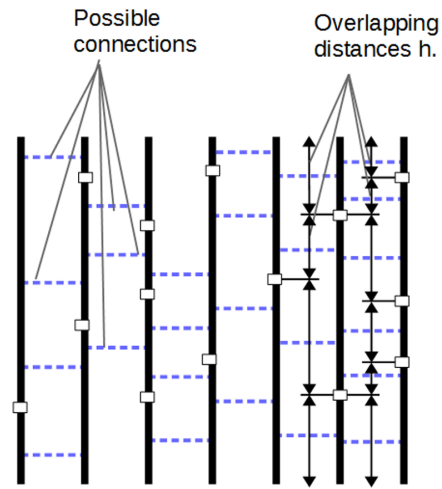}%
 \caption{\label{fig:carbon_model} Redefined graph. The overlap between two adjacent tunes is shown by intervals between the arrows.} %
\end{figure}

By definition, different segments (tubes) of the same line are not connected to
each other. Only tubes lying on adjacent lines may be connected, if
one or more open bonds between them are established. A single open bond is
sufficient to connect two tubes.  This allows one to calculate a
connection probability between two adjacent tubes in terms of
their relative position as follows. Two tubes lying on adjacent
lines may only be connected if there is a nonzero overlap $h$
between their vertical positions as shown in Fig.
\ref{fig:carbon_model}. The probability that two such tubes have
$k$ open bonds is given by the Poisson distribution with
parameter $\mu h$. That is,
\begin{equation}
 P(k)  = \frac{e^{-\mu  h}(\mu  h)^k}{k!}
\end{equation}
Tubes are disconnected ($k=0$) with probability $P(0) = e^{-\mu
h}$ and thus they are connected with probability $\pbond = 1-e^{-\mu h}$.

\begin{figure}[htp]
        \includegraphics[width=1\columnwidth]{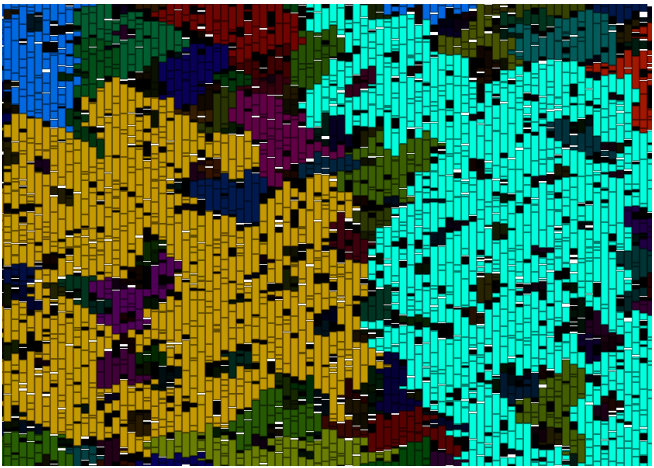}%
        \caption{\label{fig:2Dpic} A sample realization of tube-based model with clusters of connected tubes marked by a single color: the brighter the color the larger the cluster size. }
\end{figure}

A sample realization of the two dimensional model is presented in Fig. \ref{fig:2Dpic}. Groups of connected tubes form clusters marked in Fig. \ref{fig:2Dpic} by a single color.

\section{The algorithm}
\label{algorithm}
\subsection{Percolation threshold}
For any percolation model on a square lattice $L\times×L$ one
defines the crossing probability $\Pi(p,L)$  as the probability
that there is an open connection between the left
boundary and the right boundary. The crossing
probability depends on the size of the lattice and on the model
parameter $p$. In the limit $L \to \infty$, $\Pi$ converges
to $0$ for $p < \pc$ and to $1$ when $p > \pc$.  The critical
value $\pc$ is called the percolation threshold or the critical
point, and depends on the type of lattice (e.g. square,
triangular, etc.~\cite{Stauffer}).
For a finite lattice, the transition is not sharp and many approximations of the critical point are used. Examples are the point $p_\text{c1}$, where the crossing probability is equal to 0.5 \cite{grimmett2010percolation,Stauffer}, the point where the slope
of $\Pi$ (as a function of $p$) is largest, or, as used in this
paper,
\begin{equation}
\pav(L) = \int p \tfrac{d\Pi}{dp}(p,L) dp.
\label{int2}
\end{equation}
The term $\frac{d\Pi}{dp}(p,L)dp$ can be interpreted as the
probability that the graph begins to percolate for a value of the
model parameter in the interval $(p,p+dp)$. Thus $\pav$ is the
expected value of $p$ at the onset of percolation \cite{Stauffer}.
Similarly, a measure of the width of the transition region can be
defined as the variance
\begin{equation}
\Delta^2(L) = \int (p-\pav)^2 ~\tfrac{d\Pi}{dp}(p,L) dp
\label{var}
\end{equation}
These quantities satisfy the scaling relations \cite{Stauffer}:
\begin{subequations}
\begin{align}
\pav-\pc \propto L^{-\tfrac{1}{\nu}} \label{scal1}\\
\Delta \propto L^{-\tfrac{1}{\nu}} \label{scal2}
\end{align}
\end{subequations}
where $\nu$ is the (universal) critical length exponent. For
additional discussion see also
\cite{Berlyand1997,Berlyand_Wehr_1995}. This leads to an
asymptotic linear relation between $\pav(L)$ and $\Delta(L)$:
\begin{equation}
\pav = a\Delta + \pc,
\label{eq:linear}
\end{equation}
where $a$ is a proportionality constant.
Equation \ref{eq:linear} provides a simple method of extrapolating results obtained for finite lattices to the infinite one.
\subsection{Algorithms for the homogeneous lattice}
For the simplest example of an algorithm computing the critical density, consider the bond percolation model on a regular, homogeneous lattice.  We assign to each bond $i$ a random number $r_i$, sampled from the uniform distribution on the interval $[0,1]$.  To simulate a realization with density $p$, we open the bonds for which $r_i \leq p$.  We then check for existence of an open connection between the opposite sides of the lattice. Applying this with different $p$ (for the same realization of the $r_i$), we approximate $p_{\hbox{con}}$ as the value of $p$ at which the connection first forms for a given realization. The consecutive values of $p$ are selected as in the binary search algorithm. Repeating the whole procedure many times for different sets of random numbers $r_i$, we obtain a set of values $p_{\hbox{\it con}}$. This allows us to estimate $p_{\hbox{\it av}}$ by the empirical mean value of $p_{\hbox{\it con}}$ and $\Delta$ as its empirical variance. 
Such procedure is the basis of many more advanced methods of computing the percolation threshold, such as Hoshen-Kopelman \cite{hoshen1976percolation} and Leath-Alexandrowicz \cite{alexandrowicz1980critically,leath1976cluster} algorithms.
We propose to follow a different approach, which is a modification of the Newman-Ziff algorithm \cite{Ziff2001}, computing the value of $p$ at which an open connection appears for a given realization in a single run. Unlike in the original Newman-Ziff approach which used `micro-canonical ensemble' \cite{Ziff_2000,Ziff2001}, we use `canonical ensemble'. The main advantage of the modified approach is its applicability to more general graphs, where probabilities vary from bond to bond. We note that the transformation between `micro-canonical' and `canonical' ensemble representations is complicated and impractical in this generality. From the point of view of computing percolation threshold on homogeneous lattices both algorithms are equivalent, as explained in detail below. 

The idea of the so-called Rising Water algorithm, inspired by a remark in 
\cite{grimmett2010percolation} is again to assign a random 
number to every bond, as described above. To determine the value of $p$ at which percolation sets in, consecutive bonds $i$ are open in the order of the increasing $r_i$.  Assuming that random numbers assigned to
different bonds are different, at each stage we obtain the same
graph as when using the simplest method described above with
$p=r_i$. The algorithm stops when a connection linking a fixed
pair of opposite sides of the square is established. The estimate
of $\pcon$ is equal to the value $r_i$ of the last added bond.
The results of applying the two algorithms are identical. Indeed, the
first algorithm applied with $p\geq \pcon$, where $\pcon$ is a
result of the Rising Water algorithm, and the same
sequence of $r_i$ will find a connection.  On the other hand, for
($p\leq \pcon$) no connection will be found, which shows that the
two algorithms indeed yield the same result.

\subsection{Extension of the algorithm}
\label{sec:extension}

In case of the general model studied here, in which probabilities of
connections depend on both the geometry and the model parameter,
both Newman-Ziff and Rising Water algorithms need further modifications. 

In the simplest algorithm applied to the tube-based model we have to generate random values $r_i$ for all pairs of adjacent tubes and connect a pair of tubes with an overlap $h_i$ when the following condition is fulfilled:
\begin{equation}
\label{prob1}
 r_i \ge e^{-\mu h}.
\end{equation}

To define an extension of the Rising Water Algorithm we have to compute, for every bond $i$, the smallest value of model parameter $mu$ for which the equation \ref{prob1} is fulfilled. We denote this value $\mu_i$, thus
\begin{equation}
\label{prob2}
\mu_i = -\exp{r_i}/h_i.
\end{equation} 
When $\mu<\mu_i$ i-th bond is closed, and when $\mu\ge\mu_i $ i-th bond is open. 

Then, as in the homogeneous case we sort the set of $\mu_i$ in the increasing order and we open the bonds in the graph in this order. We estimate the critical value of parameter called $\mu_\text{con}$ by the first value of $\mu_i$ at which a connection between two fixed opposite sides of a square forms.

The algorithm introduced by Newman and Ziff and the extended
algorithm proposed in this paper are summarized in Table
\ref{tab:comparison}.
\begin{table*}
\center
\caption{Comparison of the Newman-Ziff algorithm with an extended algorithm.}
\footnotesize
\begin{tabular}{p{8.5cm}|p{8.5cm}}
\hline
Newman-Ziff                                               &     extended algorithm              \\
\hline
1. create a table $Q[1:N]$ to store statistic
&
1. for a given set of values of model parameter $p^s_l$ (where $l=1,2,...$) create a table $Q[...]$ to store statistic
 \\
2. run K times for k=1:K & 2. run K times for k=1:K\\
a) create a list of all bonds
& a) create a list of all bonds\\
b) generate a permutation of connections: $j_i$ means that j-th bond will be added in i-th step
&  b) assign a random number $r_i$ to every connection and compute value of model parameter $p_i$ ($\mu_i$ from Eqn. \ref{prob2}) for which we add the bond. Sort connections in order of increasing $p_i$. Let $j_i$ denote a sorting permutation \\
c) initialize the list of clusters so that each site is an a cluster of exactly one site
& c) initialize the list of clusters so that each site is an a cluster of exactly one site\\
d) for i=1:N do & d) for i=1:N do  \\
- look at bond $j_i$ connecting sites $a$ and $b$. If these sites belong to different clusters A and B, merge both clusters
& - look at bond $j_i$ connecting sites $a$ and $b$. If these sites belong to different clusters A and B, merge both clusters \\
- check for spanning: for the first occurrence save iteration number $i$ as $i_k$
& - check for spanning, for the first occurrence save $p_i$ number as $p_{\textrm{con},k}$ \\
- refresh the statistics in merged cluster and table $Q[i]$
& - refresh the statistics in merged cluster and if for any $i$, $p_{i-1} \leq p^s_l<p_i$, update the statistics $Q[p^s_l]$ \\
3. compute the percolation threshold using the values of $i_k$
& 3. compute the percolation threshold $\hat{p}_\text{av}$ and its variance $\hat{\Delta}_\text{av}$ using $p_{\text{con},k}$ as follows: $\hat{p}_\text{av} = \frac{1}{K} \sum_{k=1}^{K} p_{\text{con},k}$ and
              $\hat{\Delta}_\text{av} = \sqrt{ \frac{1}{K-1} \sum_{k=1}^{K} (p_{\text{con},k}- \hat{p}_\text{av})^2 }$  \\
4. compute the transformation from microcanonical $Q[n]$ to canonical $Q(p)$ using the following formula 
\begin{equation}
Q(p) = \sum_{n=0}^N {{N}\choose{n}} p^n (1-p)^{N-n}Q[n]
\label{eq:tranformation}
\end{equation}
& \\
\hline
\end{tabular}
\label{tab:comparison}
\end{table*}

The important parts of these algorithms are two main operations:
\begin{itemize}
\item  finding the cluster containing a given site;
\item connecting two clusters.
\end{itemize}

To make these operations efficient Newman and Ziff have proposed
to represent the connections within a graph by a so-called ``union-find'' (or ``disjoint-set'') data
structure \cite{Cormen}. It stores information about connections
in the form of trees where every site points either to another
site from the same cluster, or to itself.  The element pointing 
to itself, is the root of the tree and provides the
cluster's identification. To find the cluster containing a given site, 
we follow the path indicated by the pointers until we reach
the root. If for two sites we get the same root, both sites belong
to the same cluster. To connect two different clusters we add a
pointer between their roots. Two main modifications are commonly
used. The first one is to always point from the smaller tree to
the bigger one (``balancing''). It requires storing the information
about each cluster's size.  The second is called path compression:
having found the root of an element's cluster, we re-track the
path from the element to the root again, changing the parent of
each site along the way to the root.  Using such union-find data
structure makes operations of adding an edge and checking
whether two sites belong to the same cluster very fast. 

Beside pointer to the parent and size of the subtree, one can store additional information in
each site's record, such as moment of inertia, position, or the
information about the cluster's connection to boundaries. The last
one is a simple way to check for whether the opposite parts of the boundary are connected.  

The position of a site can be used to check whether the cluster is wrapped around the torus
\cite{Ziff_2000,Ziff2001}.

The amortized computational cost of using it  is proportional to the inverse Ackermann function and thus it can be
considered as a small constant for practical purposes \cite{Cormen}.

\subsection{Percolation statistics}
In contrast to older approaches,  the important novelty of
Newman-Ziff algorithm \cite{Ziff_2000,Ziff2001}, as shown in Table \ref{tab:comparison} (step 4),  is its ability to
simultaneously calculate a model characteristic of a given
configuration for different values of the model parameter $p$.
While standard methods need $K$ runs of the algorithm to compute $K$ values of a model characteristic for a given set of model parameters $p_k$ ($k=1,...K$), in our approach, as in that of Newman and Ziff,  all values values are obtained simultaneously in a single run. Both methods can
obtain many important characteristics of the model, for example 
average cluster size, average moment of inertia and so on,
with constant computational cost in every run of the algorithm.
Other parameters like histogram of cluster-size distribution with $B$ bins can be calculated with an additional cost proportional to the number of points in the realization ($N$) and to the number of bins. 
Let us consider a quantity $Q$. According to the Newman-Ziff algorithm we calculate $Q[i]$ which is a value of $Q$ after adding the i-th bond. The values $Q[i]$ are then averaged over $K$ different realizations, where the value of $K$ depends on the required accuracy. As the next step we transform the result to
the canonical value $Q(p)$ using Eqn. \ref{eq:tranformation}. 
In  the Rising Water algorithm we calculate $Q[j]$, the values of $Q$ for a chosen collection of values of model parameters $p_j$ and take the $Q[j]$ obtained in the last step of the algorithm (described in Table \ref{tab:comparison} as a step 4) for which we had $p<p_i$. 
In our method the possibility of effectively
achieving statistics is related to the operation on clusters.
Efficiency of our algorithm relies on fast updates of $Q$, using operations on clusters
rather than having to run through the whole graph at each step.

 For example, we consider the cluster size. The size of cluster $C$
($s_C$) obtained as a union of two clusters $A$ and $B$ is equal
to:
\begin{equation}
s_C = s_A + s_B
\end{equation}
similarly for the calculation of the moment of inertia for
clusters we use the stored quantities:  sizes of clusters $s_i$, masses of
clusters $m_i$, centers of mass $r_i$ and previous moments of
inertia $I_i$. For unions of clusters we obtain:
\begin{subequations}
\begin{align}
m_c = m_a + m_b
\label{mominert}\\
r_c =  \frac{ r_a m_a + r_b m_b}{m_c}\\
I_c = I_a + I_b + (r_a-r_c)^2*m_a + (r_b-r_c)^2*m_b
\label{stainerlarw}
\end{align}
\end{subequations}
Note that Eqn. \ref{stainerlarw} is the parallel axis theorem (Steiner law).
In our method, if we store in memory information about the clusters, all these operations have only a constant cost per operation. For example to get a mean value of the moment of inertia we additionally store in memory the sum of the moments of inertia of the clusters and update this sum.

\subsection{Critical exponents}
When the percolation threshold $\pc$ is computed, a postprocessing
algorithm gathers statistics about the distribution of clusters
(including the size of the largest cluster, cluster-size moments,
cluster-volume moments). These statistics are determined for $p$
in a vicinity of $\pc$. This allows computing several critical
exponents of the model. In particular the cluster-size
distribution near the percolation threshold allows to compute the
Fisher exponent $\tau$.  The $\beta$ exponent is computed from the
size of the maximal cluster. From data acquired in the algorithm outlined in Sec. \ref{sec:extension} exponent $\nu$ in Eqn. \ref{scal2}can be computed using the scaling relation (Eqn. \ref{scal2}).

\section{Results in the two-dimensional case}
\label{res2D}
\subsection{Percolation threshold}
\begin{figure}[htp]
\centering
\includegraphics[width=1\columnwidth]{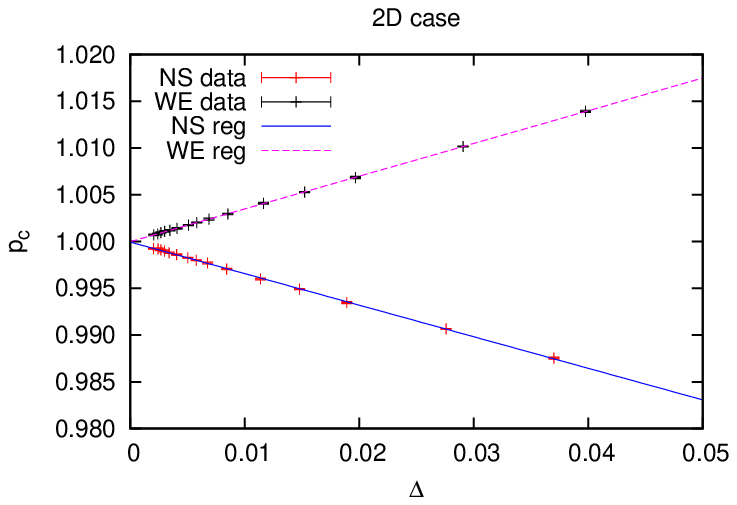}\\
a)\\
\includegraphics[width=1\columnwidth]{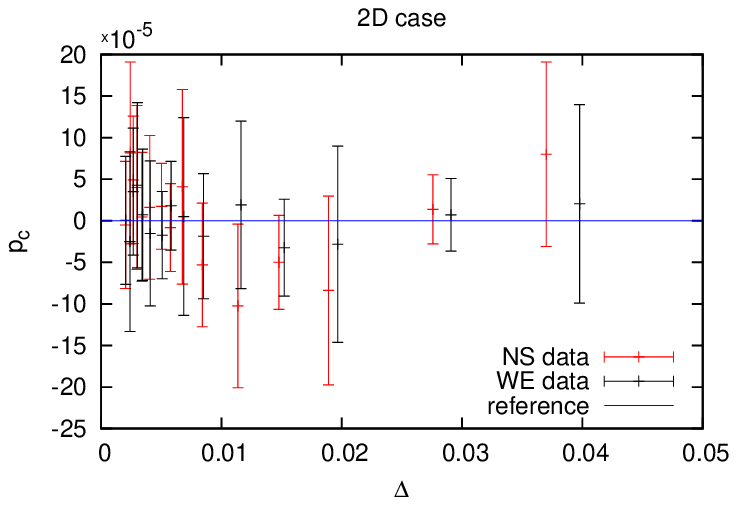}\\
b)\\
        \caption{\label{fig:2D_plot} The percolation threshold computed by studying top-to-bottom (green points) and left-to-right (red points) connections. In b) the differences between data points and the fitted line are shown.}
\end{figure}

The simulation was run for several square lattices with size ranging
from $L=200$ to $L=10000$. The estimators of $\pav$ and $\Delta$
were acquired for mutually perpendicular directions, denoted by NS (top
to bottom) and WE (left to right). The percolation threshold for
the infinite lattice ($L \rightarrow \infty$) was computed by fitting the data to the scaling properties described by Eqn. \ref{eq:linear} as presented in Fig. \ref{fig:2D_plot}. 
The results for the infinite lattice based on the intercept of the fitted linear function are the following:
\begin{subequations}
\begin{align}
p_\text{c~NS} &= 0.99999 \pm 2.5\times 10^{-5}
 \label{res2d1}\\
p_\text{c~WE} &= 0.99999 \pm 5.0 \times 10^{-5}
\end{align}
\label{res2d2}
\end{subequations}
It is worth noting that values $\pav$ converge to $\pc$ from
both directions, as presented in Fig. \ref{fig:2D_plot}. The obtained value of $\pc$ equal $1$ is clearly model-specific,
as discussed in Section \ref{anal}.
\subsection{Duality and exact analytic result}
\label{anal} 

\begin{figure}[htp]
\includegraphics[width=\columnwidth]{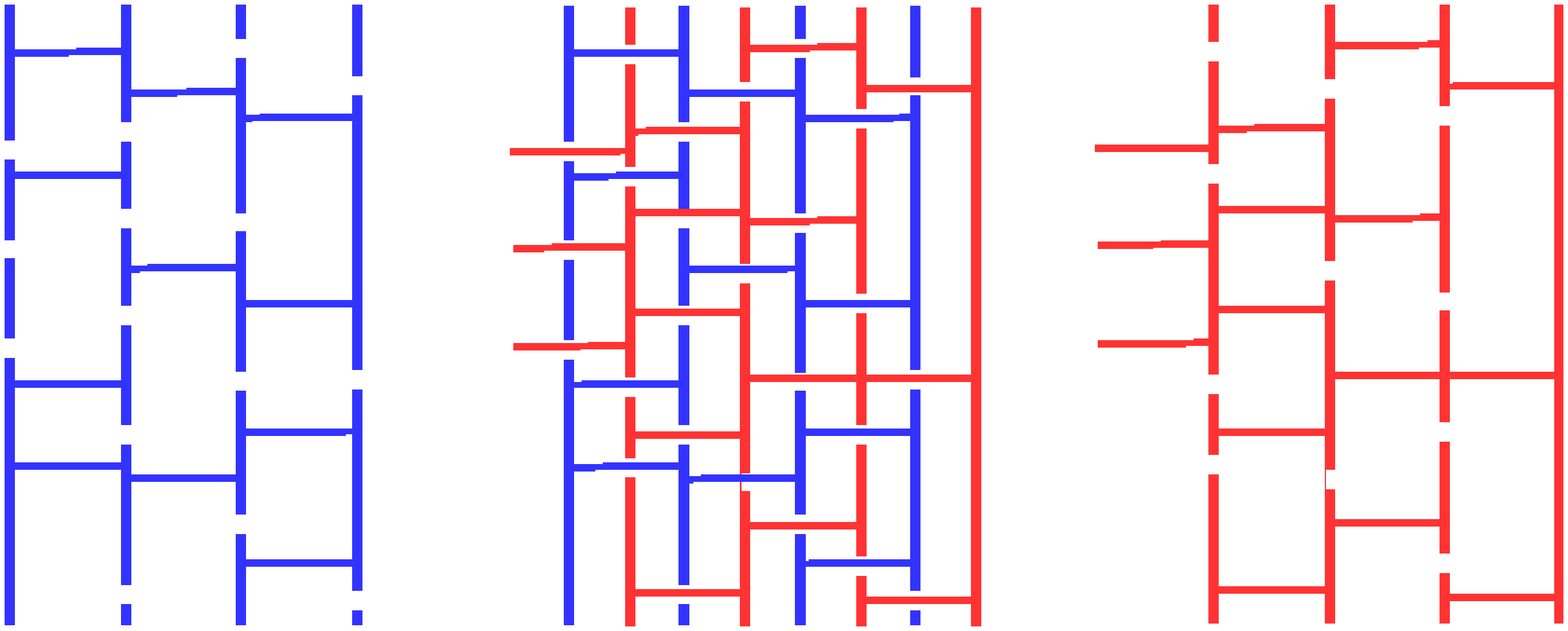}\\
a) original graph \hspace{0.1\columnwidth}     b) construction  \hspace{0.1\columnwidth}   c) dual graph\\
\caption{Construction of dual graph from the original one. 
\label{fig:dual}}%
\end{figure}

We consider a realization of the two dimensional graph defined by
$\{L, n, \mu_1, \mu_2\}$ presented in Fig. \ref{fig:dual}a. 
We define the graph dual to the initial one according to the following procedure:
\begin{itemize}
\item dual lines are introduced, each line is placed between two existing lines;
\item dual lines are divided into tubes (dual tubes) by the bonds of initial graph (vertical segments marked in Fig. \ref{fig:dual}b);
\item at the positions on breaks between initial tubes the dual bonds connecting  dual tubes are introduced (horizontal lines marked in Fig. \ref{fig:dual}b).
\end{itemize}

The two graphs, initial and dual, are shown in Fig \ref{fig:dual}a and c.

New tubes and bonds are generated in the same way as the original
ones, with the two Poisson process parameters interchanged.
Notice that the two graphs have no intersections. We either have a
connection from top to bottom, using tubes and bonds of the original 
graph, or we can
draw a line through the empty spaces and breaks between the tubes
from left to right, that does not cross any bonds or tubes. In the latter case,  there is a connection from left to right in the dual
graph.  Similarly, exactly one of the two alternatives occurs:  either there is a connection from left to right
by bonds and tubes of the original graph, or there is a connection from top to bottom in the
dual graph--- an unbroken path through empty spaces. A given
realization starts to percolate when the dual graph stops
percolating, so $\pav = \pav^{\text{dual}}$ for a pair of
dual graphs.

We know that the percolation threshold  in the limit $n=L \rightarrow
\infty$ depends only on the ratio $\mu_2/\mu_1$. Increasing
$\mu_1$ results in more (shorter) tubes and thus makes percolation
more difficult, while increasing $\mu_2$ makes for more
connections between tubes, which facilitates it.  Together with
the duality described above, this indicates that ${\mu_2 \over \mu_1} = 1$, i.e. $\mu_c =1$ is the percolation threshold, thus
explaining the numerical result (\ref{res2d1}) and (\ref{res2d2}), and
giving further support to our method.  We emphasize that a
rigorous proof that the critical value of $\mu$ equals $1$
requires a more careful argument.  The first result of this type
(for the square lattice) was proven in \cite{Kesten1980}. Simpler
arguments developed later can be found in
\cite{grimmett2010percolation}. They can be adapted to cover the
present case as well.

\subsection{Critical exponents}

Based on the scaling law (Eqn. \ref{scal1} and \ref{scal2}) we
obtain the correlation length exponent : $\nu = 1.345 \pm 0.009$. The
exact value is known to be $4/3$. 

\begin{figure}[htp]
        \includegraphics[width=1\columnwidth]{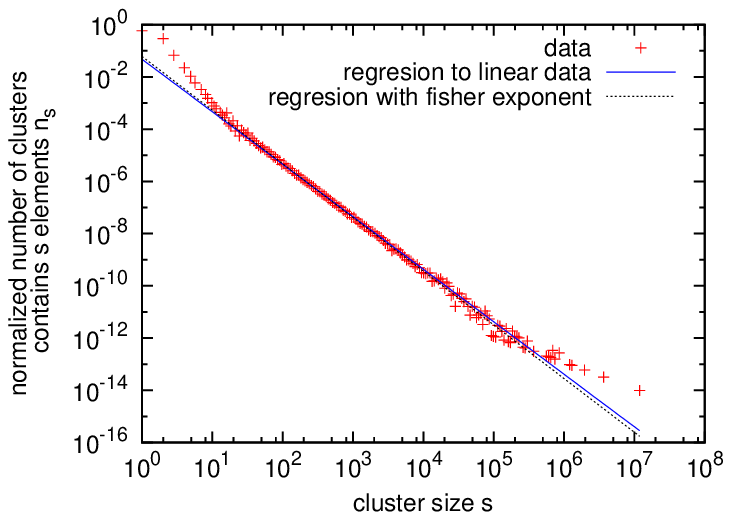}\\
        a) Normalized number of clusters of size $s$.\\
        \includegraphics[width=1\columnwidth]{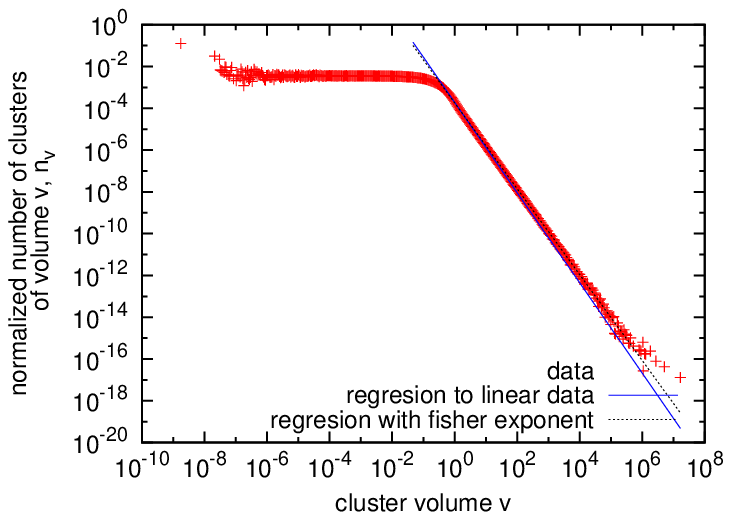}\\
        b) Normalized number of clusters of volume $v$.\\
        \caption{\label{fig:hist} Number of clusters: a) $n_s$ - number of clusters of size $s$ per one site. b) $n_v$ - number of clusters of volume $v$ per unit volume. Data from 2D grid $15000\times 15000$. 
}
\end{figure}

We determined two characteristics of the clusters: the first one, presented in Fig. \ref{fig:hist} a),
based on size of clusters and the second one, presented in Fig.
\ref{fig:hist} b), based on volume of clusters. 

The Fisher exponent $\tau$, is determined based on cluster size distribution presented in Fig. \ref{fig:hist} a) as $\tau =
2.046\pm0.023$. The exact value is $187/91 \approx 2.054$ \cite{Stauffer}.
The agreement of the results with the known values of critical exponents supports the validity of the algorithm. 

Moreover, we show that the slopes of lines fitted in Figs. \ref{fig:hist} a) and b) are the same, thus the Fisher exponent determined based on cluster size distribution and the exponent which based on cluster volume distribution are also the same.
This observation confirms the duality relation of percolation models on a given and dual graphs, discussed in Section \ref{anal}. 

\section{Results in three dimensions}
\label{sec:3D}
\begin{figure}[htp]
        \includegraphics[width=1\columnwidth]{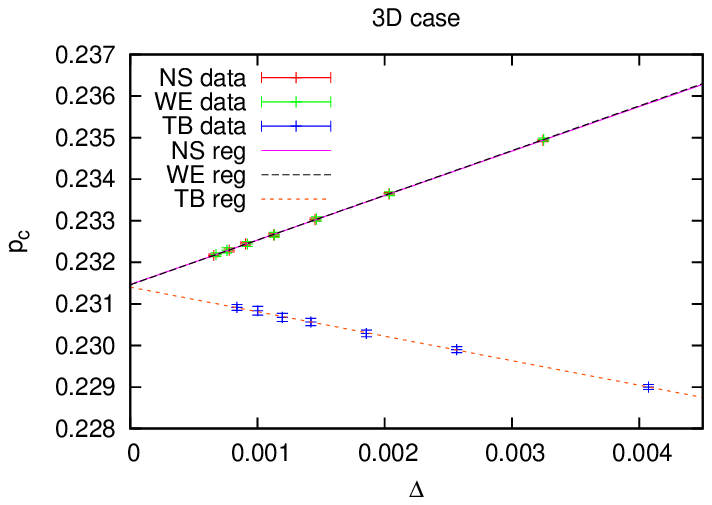}\\
				a)\\
        \includegraphics[width=1\columnwidth]{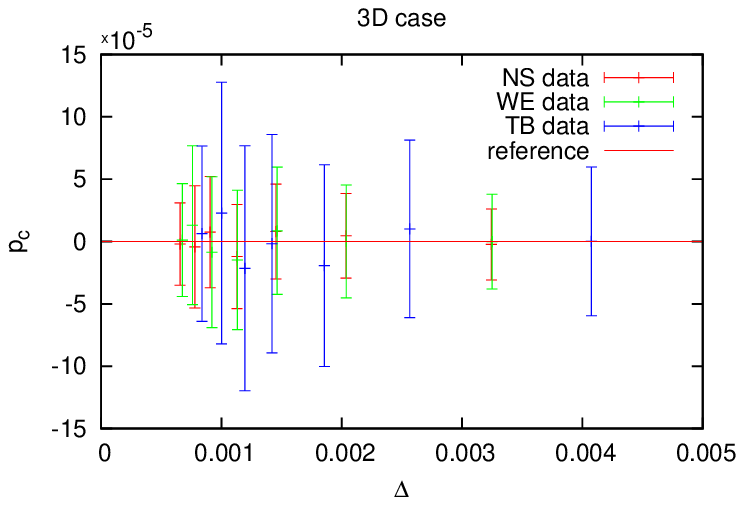}\\
				b)\\
        \caption{\label{fig:3D_plot} The  percolation threshold computed from top to bottom connections (green points) and from left to right connections (red points). In b) the differences between data points and the fitted lines are shown.}
\end{figure}

The simulation was run for cubic lattices with size ranging from $L=100$ to $L=400$.  The estimators of $\pav$ and $\Delta$ were acquired for perpendicular directions, denoted by NS, WE and TB (top to bottom).
As in (\ref{res2D}) we use scaling properties described by the Eqn. \ref{scal2} to compute the percolation threshold for infinite lattice. The results are as follows:
\begin{subequations}
\begin{align}
p_\text{c~NS} &= 0.231466 \pm 6\times 10^{-6}
\label{res3d1}\\
p_\text{c~WE} &= 0.23146 \pm 7 \times 10^{-6}
\\
p_\text{c~TB} &= 0.23140  \pm 1.2 \times 10^{-5}
\end{align}
\end{subequations}
The results obtained by fitting independently three linear functions, as presented in \ref{fig:3D_plot} a), can be improved using the following constraints: 
\begin{itemize}
\item  the lines fitted to the results perpendicular to tubes (NS and WE) have the same slope and intercept $b$; 
\item the line fitted to the results parallel to tubes (TB) have the same intercept $b$.
\end{itemize}

Thus the improved estimated value of the percolation threshold is:
\begin{subequations}
\begin{align}
\pc &= 0.231456 \pm 6\times 10^{-6}
\label{res3d2}
\end{align}
\end{subequations}
It is worth noting that, exactly as in the two-dimensional case, which
we discussed in Section (\ref{anal}), the values $\pav$
converge to $\pc$ from both directions. This is clearly visible in
Fig. \ref{fig:3D_plot}.
\section{Discussion}
\subsection{Computational cost}
The computational cost of determining an approximate value of the
critical point $\pav$ depends on the size of the lattice and on
the desired accuracy of calculation which can be expressed in terms of
standard deviation $\Delta(L)$. Analysis of this computational
complexity allows us to know what accuracy $\epsilon$ can be achieved
in a given time.
The obtained value of $\pav$ is approximated by the Monte Carlo
estimator $\hat{p}_\text{av}$, which takes into account all runs of the
algorithm:
\begin{equation}
\label{accuracy1}
\Delta\hat{p}_\text{av} \propto \frac{\sigma_{\hat{p}_\text{av}}} {\sqrt{k}}
\end{equation}
where $k$ denotes the number of repetitions of the Rising Water
algorithm. Thus $\epsilon$, the final accuracy of $\hat{p}_\text{av}$,
depends on the number of runs of the algorithm and on the variance
$\Delta(L)$ as follows:
\begin{equation}
\label{accuracy2}
\epsilon = \frac{\Delta(L)}{\sqrt{k}}
\end{equation}
From the Eqn. \ref{scal2} we know that $\Delta(L)$ depends on the size of the domain $L$.
The computational cost $c=k L^d \log \,n$ is proportional to
the number $k$ of times the Rising Water algorithm is repeated and
to the cost of a single run (of the order of $L^d \log\,n$, where
$d$ is the dimensionality of the problem). Thus the computational cost
to obtain the result with the accuracy $\epsilon$ is
\begin{equation}
\label{cost}
c = \frac{L^{d-{2 \over \nu}} \log{n}}{\epsilon^2}.
\end{equation}
The exponent $2 - {d \over \nu}$ depends on the dimension of the problem.  For
the two-dimensional case it is $1/2$, while in three dimensions it
equals approximately 0.72.
The logarithmic factor in the  expression for the computational cost (Eqn. \ref{cost}) is due to sorting of random numbers in step 2c in Section
\ref{tab:comparison}. One method to avoid this is to use so-called
bucket sort, which is a linear-time sorting algorithm using
information about data distribution \cite{Cormen}. Due to
statistical behavior of the random values $p_i$ we can create a
set of disjoint intervals that cover all possible values of $p_i$
and have approximately the same expected number of random values
$p_i$ in each interval. Let us denote this expected number of random
variables in one interval (``bucket'') by $M$. For every generated
random $p_i$ ($i=1,\ldots,N$) we can compute in constant time to
which bucket it should be assigned. When all numbers are
generated and classified, in each bucket we have a set of $M + O\left(\sqrt{M}\right)$.
numbers, and we need to sort it. The computational cost of
generating $N$ random variables and sorting $N/M$ buckets of size
$M$ is $ O ( N \log{M} ) $ and it is linear in $N$ because it is
always possible to generate enough intervals to keep M constant.
After that, the cost of running the algorithm $k$ times is
$$O(k L^d \alpha(L^d))$$
and cost of running the algorithm to get desired accuracy $\epsilon$ is
$$ O\left(\frac{L^{d-2/\nu} \alpha(L^d)}{\epsilon^2} \right)$$
Here $\alpha(L^d)$ is the Ackerman function and can be considered
constant. 
Despite better asymptotic behavior of the bucket sort, it
does not give a better performance except for very big lattices. 

\section{Conclusions}
In summary, three goals have been achieved in this work:
\begin{itemize}
\item We have defined a family of discrete-continuous
percolation models motivated by the physics of activated carbon.
These models deal with tubes of random length connected by random bonds; 
as such they should describe well situations in which complicated micro-structures
observed in activated carbon have approximately linear textures: graphene ribbons (model of Jenkins-Kawamura)
and nanotubes \cite{harris2013fullerene}. In cases the structures are neither
1D nor quasi-1D (fullerenes (Harris
model \cite{harris2003impact,marsh2006activated}), stacked
graphite \cite{jenkins1971structure}, or graphene
\cite{pawlyta2013transmission}, carbon onions
\cite{chen2006pore}), the concrete  models considered here provide
only a ``caricature'' of the real situation. Still we expect that
even in these cases they capture some qualitative aspects of the underlying
physics.
\item We have extended the standard algorithm  of Newman and Ziff \cite{Ziff2001} to handle inhomogeneous lattices. This extension is non-trivial, and we have analyzed in detail its convergence properties.
\item We applied the extended algorithm to the family of models in
question, calculating critical parameters and cluster density
distributions in two and three dimensions.
\end{itemize}
Possibilities for further studies include: i) applications of the
present models to experimental data, suggesting geometry formed by
parallel random tubes/ribbons connected randomly by bonds; ii)
development of concrete models with geometry formed by parallel
random flakes/patches connected randomly by bonds; iii)
application of the method to such models, calculation of their
properties, and direct comparison with experiments.

It is worth mentioning that the problem of quantum aspects of
the carbon activation process is also to a great extent open. This
suggests to study quantum versions of the family of the
discrete-continuous models discussed in this paper. The interplay
of discrete and continuous aspects may lead to quantitatively novel
effects. It is worth noting that such quantum disordered models can in principle be simulated, {\it quantum simulated}, by a system of ultracold atoms (see, for instance, \cite{Lewenstein2012}): an array of random length 1D Bose
condensed gases with controlled random connections between them.

\begin{acknowledgments}
This work has been partially supported by the Iuventus Plus
programme founded by the Polish Ministry of Science and Higher
Education (IP2014 024373). M.L. acknowledges Spanish MINECO Project FOQUS
(FIS2013-46768), ERC AdG OSYRIS, EU IP SIQS, EU STREP EQuaM, and
EU FETPROACT QUIC.   J.W. has been partially funded by NSF grant
DMS 131271.
\end{acknowledgments}
\newpage
\bibliographystyle{apsrev4-1}
\bibliography{paperNotes}

\begin{thebibliography}{29}%
\makeatletter
\providecommand \@ifxundefined [1]{%
 \@ifx{#1\undefined}
}%
\providecommand \@ifnum [1]{%
 \ifnum #1\expandafter \@firstoftwo
 \else \expandafter \@secondoftwo
 \fi
}%
\providecommand \@ifx [1]{%
 \ifx #1\expandafter \@firstoftwo
 \else \expandafter \@secondoftwo
 \fi
}%
\providecommand \natexlab [1]{#1}%
\providecommand \enquote  [1]{``#1''}%
\providecommand \bibnamefont  [1]{#1}%
\providecommand \bibfnamefont [1]{#1}%
\providecommand \citenamefont [1]{#1}%
\providecommand \href@noop [0]{\@secondoftwo}%
\providecommand \href [0]{\begingroup \@sanitize@url \@href}%
\providecommand \@href[1]{\@@startlink{#1}\@@href}%
\providecommand \@@href[1]{\endgroup#1\@@endlink}%
\providecommand \@sanitize@url [0]{\catcode `\\12\catcode `\$12\catcode
  `\&12\catcode `\#12\catcode `\^12\catcode `\_12\catcode `\%12\relax}%
\providecommand \@@startlink[1]{}%
\providecommand \@@endlink[0]{}%
\providecommand \url  [0]{\begingroup\@sanitize@url \@url }%
\providecommand \@url [1]{\endgroup\@href {#1}{\urlprefix }}%
\providecommand \urlprefix  [0]{URL }%
\providecommand \Eprint [0]{\href }%
\providecommand \doibase [0]{http://dx.doi.org/}%
\providecommand \selectlanguage [0]{\@gobble}%
\providecommand \bibinfo  [0]{\@secondoftwo}%
\providecommand \bibfield  [0]{\@secondoftwo}%
\providecommand \translation [1]{[#1]}%
\providecommand \BibitemOpen [0]{}%
\providecommand \bibitemStop [0]{}%
\providecommand \bibitemNoStop [0]{.\EOS\space}%
\providecommand \EOS [0]{\spacefactor3000\relax}%
\providecommand \BibitemShut  [1]{\csname bibitem#1\endcsname}%
\let\auto@bib@innerbib\@empty
\bibitem [{\citenamefont {Grimmett}(2010)}]{grimmett2010percolation}%
  \BibitemOpen
  \bibfield  {author} {\bibinfo {author} {\bibfnamefont {G.~R.}\ \bibnamefont
  {Grimmett}},\ }\href@noop {} {\emph {\bibinfo {title} {Percolation
  (Grundlehren der mathematischen Wissenschaften)}}}\ (\bibinfo  {publisher}
  {Springer: Berlin, Germany},\ \bibinfo {year} {2010})\BibitemShut {NoStop}%
\bibitem [{\citenamefont {Aharony}\ and\ \citenamefont
  {Stauffer}(2003)}]{Stauffer}%
  \BibitemOpen
  \bibfield  {author} {\bibinfo {author} {\bibfnamefont {A.}~\bibnamefont
  {Aharony}}\ and\ \bibinfo {author} {\bibfnamefont {D.}~\bibnamefont
  {Stauffer}},\ }\href@noop {} {\emph {\bibinfo {title} {Introduction to
  percolation theory}}}\ (\bibinfo  {publisher} {Taylor \& Francis},\ \bibinfo
  {address} {United Kingdom},\ \bibinfo {year} {2003})\BibitemShut {NoStop}%
\bibitem [{\citenamefont {Meester}\ and\ \citenamefont
  {Roy}(1996)}]{meester1996continuum}%
  \BibitemOpen
  \bibfield  {author} {\bibinfo {author} {\bibfnamefont {R.}~\bibnamefont
  {Meester}}\ and\ \bibinfo {author} {\bibfnamefont {R.}~\bibnamefont {Roy}},\
  }\href@noop {} {\emph {\bibinfo {title} {Continuum percolation}}},\ \bibinfo
  {series} {Cambridge Tracts in Mathematics}, Vol.\ \bibinfo {volume} {119}\
  (\bibinfo  {publisher} {Cambridge University Press, Cambridge},\ \bibinfo
  {year} {1996})\BibitemShut {NoStop}%
\bibitem [{\citenamefont {Wu}\ \emph {et~al.}(2006)\citenamefont {Wu},
  \citenamefont {Braunstein}, \citenamefont {Havlin},\ and\ \citenamefont
  {Stanley}}]{PhysRevLett.96.148702}%
  \BibitemOpen
  \bibfield  {author} {\bibinfo {author} {\bibfnamefont {Z.}~\bibnamefont
  {Wu}}, \bibinfo {author} {\bibfnamefont {L.~A.}\ \bibnamefont {Braunstein}},
  \bibinfo {author} {\bibfnamefont {S.}~\bibnamefont {Havlin}}, \ and\ \bibinfo
  {author} {\bibfnamefont {H.~E.}\ \bibnamefont {Stanley}},\ }\href {\doibase
  10.1103/PhysRevLett.96.148702} {\bibfield  {journal} {\bibinfo  {journal}
  {Phys. Rev. Lett.}\ }\textbf {\bibinfo {volume} {96}},\ \bibinfo {pages}
  {148702} (\bibinfo {year} {2006})}\BibitemShut {NoStop}%
\bibitem [{\citenamefont {Li}\ \emph {et~al.}(2015)\citenamefont {Li},
  \citenamefont {Fu}, \citenamefont {Wang}, \citenamefont {Lu}, \citenamefont
  {Berezin}, \citenamefont {Stanley},\ and\ \citenamefont
  {Havlin}}]{Li20012015}%
  \BibitemOpen
  \bibfield  {author} {\bibinfo {author} {\bibfnamefont {D.}~\bibnamefont
  {Li}}, \bibinfo {author} {\bibfnamefont {B.}~\bibnamefont {Fu}}, \bibinfo
  {author} {\bibfnamefont {Y.}~\bibnamefont {Wang}}, \bibinfo {author}
  {\bibfnamefont {G.}~\bibnamefont {Lu}}, \bibinfo {author} {\bibfnamefont
  {Y.}~\bibnamefont {Berezin}}, \bibinfo {author} {\bibfnamefont {H.~E.}\
  \bibnamefont {Stanley}}, \ and\ \bibinfo {author} {\bibfnamefont
  {S.}~\bibnamefont {Havlin}},\ }\href {\doibase 10.1073/pnas.1419185112}
  {\bibfield  {journal} {\bibinfo  {journal} {Proceedings of the National
  Academy of Sciences}\ }\textbf {\bibinfo {volume} {112}},\ \bibinfo {pages}
  {669} (\bibinfo {year} {2015})},\ \Eprint
  {http://arxiv.org/abs/http://www.pnas.org/content/112/3/669.full.pdf}
  {http://www.pnas.org/content/112/3/669.full.pdf} \BibitemShut {NoStop}%
\bibitem [{\citenamefont {Franceschetti}\ \emph {et~al.}(2005)\citenamefont
  {Franceschetti}, \citenamefont {Booth}, \citenamefont {Cook}, \citenamefont
  {Meester},\ and\ \citenamefont {Bruck}}]{franceschetti2005continuum}%
  \BibitemOpen
  \bibfield  {author} {\bibinfo {author} {\bibfnamefont {M.}~\bibnamefont
  {Franceschetti}}, \bibinfo {author} {\bibfnamefont {L.}~\bibnamefont
  {Booth}}, \bibinfo {author} {\bibfnamefont {M.}~\bibnamefont {Cook}},
  \bibinfo {author} {\bibfnamefont {R.}~\bibnamefont {Meester}}, \ and\
  \bibinfo {author} {\bibfnamefont {J.}~\bibnamefont {Bruck}},\ }\href@noop {}
  {\bibfield  {journal} {\bibinfo  {journal} {Journal of Statistical Physics}\
  }\textbf {\bibinfo {volume} {118}},\ \bibinfo {pages} {721} (\bibinfo {year}
  {2005})}\BibitemShut {NoStop}%
\bibitem [{\citenamefont {Booth}\ \emph {et~al.}(2003)\citenamefont {Booth},
  \citenamefont {Bruck}, \citenamefont {Franceschetti},\ and\ \citenamefont
  {Meester}}]{Discs_2003}%
  \BibitemOpen
  \bibfield  {author} {\bibinfo {author} {\bibfnamefont {L.}~\bibnamefont
  {Booth}}, \bibinfo {author} {\bibfnamefont {J.}~\bibnamefont {Bruck}},
  \bibinfo {author} {\bibfnamefont {M.}~\bibnamefont {Franceschetti}}, \ and\
  \bibinfo {author} {\bibfnamefont {R.}~\bibnamefont {Meester}},\ }\href
  {http://www.jstor.org/stable/1193166} {\bibfield  {journal} {\bibinfo
  {journal} {The Annals of Applied Probability}\ }\textbf {\bibinfo {volume}
  {13}},\ \bibinfo {pages} {pp. 722} (\bibinfo {year} {2003})}\BibitemShut
  {NoStop}%
\bibitem [{\citenamefont {Furmaniak}\ \emph {et~al.}(2014)\citenamefont
  {Furmaniak}, \citenamefont {Terzyk}, \citenamefont {Gauden}, \citenamefont
  {Kowalczyk},\ and\ \citenamefont {Harris}}]{furmaniak2014folding}%
  \BibitemOpen
  \bibfield  {author} {\bibinfo {author} {\bibfnamefont {S.}~\bibnamefont
  {Furmaniak}}, \bibinfo {author} {\bibfnamefont {A.~P.}\ \bibnamefont
  {Terzyk}}, \bibinfo {author} {\bibfnamefont {P.~A.}\ \bibnamefont {Gauden}},
  \bibinfo {author} {\bibfnamefont {P.}~\bibnamefont {Kowalczyk}}, \ and\
  \bibinfo {author} {\bibfnamefont {P.~J.}\ \bibnamefont {Harris}},\
  }\href@noop {} {\bibfield  {journal} {\bibinfo  {journal} {Journal of
  Physics: Condensed Matter}\ }\textbf {\bibinfo {volume} {26}},\ \bibinfo
  {pages} {485006} (\bibinfo {year} {2014})}\BibitemShut {NoStop}%
\bibitem [{\citenamefont {Furmaniak}\ \emph {et~al.}(2012)\citenamefont
  {Furmaniak}, \citenamefont {Terzyk}, \citenamefont {Gauden},\ and\
  \citenamefont {Kowalczyk}}]{furmaniak2012simulation}%
  \BibitemOpen
  \bibfield  {author} {\bibinfo {author} {\bibfnamefont {S.}~\bibnamefont
  {Furmaniak}}, \bibinfo {author} {\bibfnamefont {A.~P.}\ \bibnamefont
  {Terzyk}}, \bibinfo {author} {\bibfnamefont {P.~A.}\ \bibnamefont {Gauden}},
  \ and\ \bibinfo {author} {\bibfnamefont {P.}~\bibnamefont {Kowalczyk}},\
  }\href@noop {} {\bibfield  {journal} {\bibinfo  {journal} {Microporous and
  Mesoporous Materials}\ }\textbf {\bibinfo {volume} {154}},\ \bibinfo {pages}
  {51} (\bibinfo {year} {2012})}\BibitemShut {NoStop}%
\bibitem [{\citenamefont {Newman}\ and\ \citenamefont {Ziff}(2001)}]{Ziff2001}%
  \BibitemOpen
  \bibfield  {author} {\bibinfo {author} {\bibfnamefont {M.~E.}\ \bibnamefont
  {Newman}}\ and\ \bibinfo {author} {\bibfnamefont {R.~M.}\ \bibnamefont
  {Ziff}},\ }\href@noop {} {\bibfield  {journal} {\bibinfo  {journal} {Physical
  Review E}\ }\textbf {\bibinfo {volume} {64}},\ \bibinfo {pages} {016706}
  (\bibinfo {year} {2001})}\BibitemShut {NoStop}%
\bibitem [{\citenamefont {Feng}\ and\ \citenamefont {Bhatia}(2000)}]{Feng2000}%
  \BibitemOpen
  \bibfield  {author} {\bibinfo {author} {\bibfnamefont {B.}~\bibnamefont
  {Feng}}\ and\ \bibinfo {author} {\bibfnamefont {S.~K.}\ \bibnamefont
  {Bhatia}},\ }\href@noop {} {\bibfield  {journal} {\bibinfo  {journal} {Energy
  and Fuels}\ }\textbf {\bibinfo {volume} {14}},\ \bibinfo {pages} {297}
  (\bibinfo {year} {2000})}\BibitemShut {NoStop}%
\bibitem [{\citenamefont {Kwiatkowski}\ \emph {et~al.}(2014)\citenamefont
  {Kwiatkowski}, \citenamefont {Bajer}, \citenamefont {Celi\'{n}ska},
  \citenamefont {Dudy\'{n}ski}, \citenamefont {Korotko},\ and\ \citenamefont
  {Sosnowska}}]{Kwiatkowski2014125}%
  \BibitemOpen
  \bibfield  {author} {\bibinfo {author} {\bibfnamefont {K.}~\bibnamefont
  {Kwiatkowski}}, \bibinfo {author} {\bibfnamefont {K.}~\bibnamefont {Bajer}},
  \bibinfo {author} {\bibfnamefont {A.}~\bibnamefont {Celi\'{n}ska}}, \bibinfo
  {author} {\bibfnamefont {M.}~\bibnamefont {Dudy\'{n}ski}}, \bibinfo {author}
  {\bibfnamefont {J.}~\bibnamefont {Korotko}}, \ and\ \bibinfo {author}
  {\bibfnamefont {M.}~\bibnamefont {Sosnowska}},\ }\href@noop {} {\bibfield
  {journal} {\bibinfo  {journal} {Fuel}\ }\textbf {\bibinfo {volume} {132}},\
  \bibinfo {pages} {125 } (\bibinfo {year} {2014})}\BibitemShut {NoStop}%
\bibitem [{\citenamefont {Marsh}(2001)}]{marsh2001activated}%
  \BibitemOpen
  \bibfield  {author} {\bibinfo {author} {\bibfnamefont {H.}~\bibnamefont
  {Marsh}},\ }\href@noop {} {\emph {\bibinfo {title} {Activated carbon
  compendium: a collection of papers from the journal carbon 1996-2000}}}\
  (\bibinfo  {publisher} {Gulf Professional Publishing},\ \bibinfo {year}
  {2001})\BibitemShut {NoStop}%
\bibitem [{\citenamefont {Pawlyta}(2013)}]{pawlyta2013transmission}%
  \BibitemOpen
  \bibfield  {author} {\bibinfo {author} {\bibfnamefont {M.}~\bibnamefont
  {Pawlyta}},\ }\href@noop {} {\bibfield  {journal} {\bibinfo  {journal}
  {Materials Science and Engineering}\ }\textbf {\bibinfo {volume} {63}},\
  \bibinfo {pages} {58} (\bibinfo {year} {2013})}\BibitemShut {NoStop}%
\bibitem [{\citenamefont {Harris}(2003)}]{harris2003impact}%
  \BibitemOpen
  \bibfield  {author} {\bibinfo {author} {\bibfnamefont {P.~J.}\ \bibnamefont
  {Harris}},\ }\href@noop {} {\bibfield  {journal} {\bibinfo  {journal}
  {Chemistry and physics of carbon}\ }\textbf {\bibinfo {volume} {28}}
  (\bibinfo {year} {2003})}\BibitemShut {NoStop}%
\bibitem [{\citenamefont {Marsh}\ and\ \citenamefont
  {Reinoso}(2006)}]{marsh2006activated}%
  \BibitemOpen
  \bibfield  {author} {\bibinfo {author} {\bibfnamefont {H.}~\bibnamefont
  {Marsh}}\ and\ \bibinfo {author} {\bibfnamefont {F.~R.}\ \bibnamefont
  {Reinoso}},\ }\href@noop {} {\emph {\bibinfo {title} {Activated carbon}}}\
  (\bibinfo  {publisher} {Elsevier},\ \bibinfo {year} {2006})\BibitemShut
  {NoStop}%
\bibitem [{\citenamefont {Jenkins}\ and\ \citenamefont
  {Kawamura}(1971)}]{jenkins1971structure}%
  \BibitemOpen
  \bibfield  {author} {\bibinfo {author} {\bibfnamefont {G.}~\bibnamefont
  {Jenkins}}\ and\ \bibinfo {author} {\bibfnamefont {K.}~\bibnamefont
  {Kawamura}},\ }\href@noop {} {\bibfield  {journal} {\bibinfo  {journal}
  {Nature}\ }\textbf {\bibinfo {volume} {231}},\ \bibinfo {pages} {175}
  (\bibinfo {year} {1971})}\BibitemShut {NoStop}%
\bibitem [{\citenamefont {Chen}\ \emph {et~al.}(2006)\citenamefont {Chen},
  \citenamefont {Liu}, \citenamefont {Li},\ and\ \citenamefont
  {Cheng}}]{chen2006pore}%
  \BibitemOpen
  \bibfield  {author} {\bibinfo {author} {\bibfnamefont {Y.}~\bibnamefont
  {Chen}}, \bibinfo {author} {\bibfnamefont {C.}~\bibnamefont {Liu}}, \bibinfo
  {author} {\bibfnamefont {F.}~\bibnamefont {Li}}, \ and\ \bibinfo {author}
  {\bibfnamefont {H.-M.}\ \bibnamefont {Cheng}},\ }\href@noop {} {\bibfield
  {journal} {\bibinfo  {journal} {Journal of Porous Materials}\ }\textbf
  {\bibinfo {volume} {13}},\ \bibinfo {pages} {141} (\bibinfo {year}
  {2006})}\BibitemShut {NoStop}%
\bibitem [{\citenamefont {Harris}(2013)}]{harris2013fullerene}%
  \BibitemOpen
  \bibfield  {author} {\bibinfo {author} {\bibfnamefont {P.~J.}\ \bibnamefont
  {Harris}},\ }\href@noop {} {\bibfield  {journal} {\bibinfo  {journal}
  {Journal of Materials Science}\ }\textbf {\bibinfo {volume} {48}},\ \bibinfo
  {pages} {565} (\bibinfo {year} {2013})}\BibitemShut {NoStop}%
\bibitem [{\citenamefont {Wang}\ \emph {et~al.}(2014)\citenamefont {Wang},
  \citenamefont {Sun},\ and\ \citenamefont {Chen}}]{wang2014three}%
  \BibitemOpen
  \bibfield  {author} {\bibinfo {author} {\bibfnamefont {X.}~\bibnamefont
  {Wang}}, \bibinfo {author} {\bibfnamefont {G.}~\bibnamefont {Sun}}, \ and\
  \bibinfo {author} {\bibfnamefont {P.}~\bibnamefont {Chen}},\ }\href@noop {}
  {\bibfield  {journal} {\bibinfo  {journal} {Frontiers in Energy Research}\
  }\textbf {\bibinfo {volume} {2}},\ \bibinfo {pages} {33} (\bibinfo {year}
  {2014})}\BibitemShut {NoStop}%
\bibitem [{\citenamefont {Berlyand}\ and\ \citenamefont
  {Wehr}(1997)}]{Berlyand1997}%
  \BibitemOpen
  \bibfield  {author} {\bibinfo {author} {\bibfnamefont {L.}~\bibnamefont
  {Berlyand}}\ and\ \bibinfo {author} {\bibfnamefont {J.}~\bibnamefont
  {Wehr}},\ }\href {\doibase 10.1007/s002200050082} {\bibfield  {journal}
  {\bibinfo  {journal} {Communications in Mathematical Physics}\ }\textbf
  {\bibinfo {volume} {185}},\ \bibinfo {pages} {73} (\bibinfo {year}
  {1997})}\BibitemShut {NoStop}%
\bibitem [{\citenamefont {Berlyand}\ and\ \citenamefont
  {Wehr}(1995)}]{Berlyand_Wehr_1995}%
  \BibitemOpen
  \bibfield  {author} {\bibinfo {author} {\bibfnamefont {L.}~\bibnamefont
  {Berlyand}}\ and\ \bibinfo {author} {\bibfnamefont {J.}~\bibnamefont
  {Wehr}},\ }\href {http://stacks.iop.org/0305-4470/28/i=24/a=013} {\bibfield
  {journal} {\bibinfo  {journal} {Journal of Physics A: Mathematical and
  General}\ }\textbf {\bibinfo {volume} {28}},\ \bibinfo {pages} {7127}
  (\bibinfo {year} {1995})}\BibitemShut {NoStop}%
\bibitem [{\citenamefont {Hoshen}\ and\ \citenamefont
  {Kopelman}(1976)}]{hoshen1976percolation}%
  \BibitemOpen
  \bibfield  {author} {\bibinfo {author} {\bibfnamefont {J.}~\bibnamefont
  {Hoshen}}\ and\ \bibinfo {author} {\bibfnamefont {R.}~\bibnamefont
  {Kopelman}},\ }\href@noop {} {\bibfield  {journal} {\bibinfo  {journal}
  {Physical Review B}\ }\textbf {\bibinfo {volume} {14}},\ \bibinfo {pages}
  {3438} (\bibinfo {year} {1976})}\BibitemShut {NoStop}%
\bibitem [{\citenamefont {Alexandrowicz}(1980)}]{alexandrowicz1980critically}%
  \BibitemOpen
  \bibfield  {author} {\bibinfo {author} {\bibfnamefont {Z.}~\bibnamefont
  {Alexandrowicz}},\ }\href@noop {} {\bibfield  {journal} {\bibinfo  {journal}
  {Physics Letters A}\ }\textbf {\bibinfo {volume} {80}},\ \bibinfo {pages}
  {284} (\bibinfo {year} {1980})}\BibitemShut {NoStop}%
\bibitem [{\citenamefont {Leath}(1976)}]{leath1976cluster}%
  \BibitemOpen
  \bibfield  {author} {\bibinfo {author} {\bibfnamefont {P.}~\bibnamefont
  {Leath}},\ }\href@noop {} {\bibfield  {journal} {\bibinfo  {journal}
  {Physical Review B}\ }\textbf {\bibinfo {volume} {14}},\ \bibinfo {pages}
  {5046} (\bibinfo {year} {1976})}\BibitemShut {NoStop}%
\bibitem [{\citenamefont {Newman}\ and\ \citenamefont
  {Ziff}(2000)}]{Ziff_2000}%
  \BibitemOpen
  \bibfield  {author} {\bibinfo {author} {\bibfnamefont {M.~E.~J.}\
  \bibnamefont {Newman}}\ and\ \bibinfo {author} {\bibfnamefont {R.~M.}\
  \bibnamefont {Ziff}},\ }\href {\doibase 10.1103/PhysRevLett.85.4104}
  {\bibfield  {journal} {\bibinfo  {journal} {Phys. Rev. Lett.}\ }\textbf
  {\bibinfo {volume} {85}},\ \bibinfo {pages} {4104} (\bibinfo {year}
  {2000})}\BibitemShut {NoStop}%
\bibitem [{\citenamefont {Cormen}(2009)}]{Cormen}%
  \BibitemOpen
  \bibfield  {author} {\bibinfo {author} {\bibfnamefont {T.~H.}\ \bibnamefont
  {Cormen}},\ }\href@noop {} {\emph {\bibinfo {title} {Introduction to
  algorithms}}}\ (\bibinfo  {publisher} {MIT press},\ \bibinfo {year}
  {2009})\BibitemShut {NoStop}%
\bibitem [{\citenamefont {Kesten}(1980)}]{Kesten1980}%
  \BibitemOpen
  \bibfield  {author} {\bibinfo {author} {\bibfnamefont {H.}~\bibnamefont
  {Kesten}},\ }\href {\doibase 10.1007/BF01197577} {\bibfield  {journal}
  {\bibinfo  {journal} {Communications in Mathematical Physics}\ }\textbf
  {\bibinfo {volume} {74}},\ \bibinfo {pages} {41} (\bibinfo {year}
  {1980})}\BibitemShut {NoStop}%
\bibitem [{\citenamefont {Lewenstein}\ \emph {et~al.}(2012)\citenamefont
  {Lewenstein}, \citenamefont {Sanpera},\ and\ \citenamefont
  {Ahufinger}}]{Lewenstein2012}%
  \BibitemOpen
  \bibfield  {author} {\bibinfo {author} {\bibfnamefont {M.}~\bibnamefont
  {Lewenstein}}, \bibinfo {author} {\bibfnamefont {A.}~\bibnamefont {Sanpera}},
  \ and\ \bibinfo {author} {\bibfnamefont {V.}~\bibnamefont {Ahufinger}},\
  }\href@noop {} {\emph {\bibinfo {title} {Ultracold atoms in optical lattices:
  Simulating quantum many-body systems}}}\ (\bibinfo  {publisher} {Oxford
  University Press},\ \bibinfo {address} {Oxford},\ \bibinfo {year}
  {2012})\BibitemShut {NoStop}%
\end{thebibliography}%

\end{document}